\documentclass[aps,prb,preprint,12pt,floatfix]{revtex4}
\usepackage{epsfig}  

\newcommand{\gl}[1]{Eq. (\ref{#1})}
\newcommand{\gls}[2]{Eqs. (\ref{#1},\ref{#2})}
\newcommand{\glto}[2]{Eqs. (\ref{#1}) to (\ref{#2})}
\def\gtrless{\raise2.5pt\hbox{$>$}\llap{\lower2.5pt\hbox{$<$}}}
\def\gtrapprox{\raise2.5pt\hbox{$>$}\llap{\lower2.5pt\hbox{$\approx$}}}
\newcommand{\bsq}[1]{\begin{subequations}\label{#1}}
\newcommand{\esq}{\end{subequations}}
\newcommand{\beq}[1]{\begin{equation}\label{#1}}
\newcommand{\eeq}{\end{equation}}
\newcommand{\beqa}[1]{\begin{eqnarray}\label{#1}}
\newcommand{\eeqa}{\end{eqnarray}}

\newcommand{\vek}[1]{{\bf #1}} 
\newcommand{\gd}{\dot{\gamma}}
\newcommand{\kap}{\mbox{\boldmath $\kappa$}}

\renewcommand{\rho}{\varrho}
\newcommand{\eps}{\varepsilon}

\def\gtrapprox{\raise2.5pt\hbox{$>$}\llap{\lower2.5pt\hbox{$\approx$}}}
\def\lessapprox{\raise2.5pt\hbox{$<$}\llap{\lower2.5pt\hbox{$\approx$}}}

\begin{document}

\title{Schematic models for dynamic yielding of sheared colloidal glasses}
 
\author{Matthias Fuchs\footnote{Present address:
Institut Charles Sadron, 6, rue Boussingault, 67083
Strasbourg Cedex, France;\\
permanent address:
Physik-Department, Technische Universit{\"a}t M{\"u}nchen,   85747
Garching, Germany.} 
 and Michael E. Cates}

\affiliation{Department of Physics and Astronomy, The University of
Edinburgh, JCMB King's Buildings, Edinburgh EH9 3JZ, GB}

\date{\today}

\begin{abstract}
\rule{\textwidth}{1pt}
The nonlinear rheological properties of dense suspensions are discussed within
simplified models, suggested by a recent first principles approach to
the model of Brownian particles in a constant-velocity-gradient solvent flow.
Shear thinning of colloidal fluids and  dynamical yielding of
colloidal glasses arise from a competition between a slowing down of
structural relaxation, because of particle interactions, and enhanced
decorrelation of fluctuations, caused by the shear advection of 
density fluctuations. 
A mode coupling approach is developed to explore the shear-induced
suppression of  particle caging and the resulting speed-up of the
structural relaxation. 
\rule{\textwidth}{1pt}
\end{abstract}
\maketitle

\section{Introduction}

Soft materials, such as particle dispersions, exhibit a wide range of
rheological properties. While dilute colloids flow with a viscosity 
only slightly higher than that of the solvent, concentrated
dispersions behave as weak amorphous solids. For intermediate
concentrations, one generally observes,  upon increasing the external
shear rate, first, a strong decrease of the
dispersion viscosity (``shear
thinning''), and then an (often dramatic) increase of the viscosity
(``shear thickening'') \cite{larson,russel}. 

While shear-induced crystallization of particle suspensions causes a
marked decrease of the viscosity, shear thinning is not always accompanied by a
flow-induced ordering \cite{Laun92,Bender96,Strating99,Foss00}, but appears
connected more generally to a decrease of the Brownian contribution to
the stress \cite{Brady96,Bergenholtz01b}. In concentrated suspensions of
polydisperse colloidal particles the structure remains amorphous
during the application of shear but still exhibits shear thinning or yield behaviour \cite{Petekidis02,Petekidis02b}. 

Detailed light scattering studies of quiescent colloidal hard sphere
suspensions \cite{Megen91,Megen94,Beck99,Bartsch02} have identified a
slowing down of the structural relaxation as the origin of the
solid-like behavior at high concentrations. Operationally, a
transition to an amorphous solid or glass can be defined when the
structural relaxation time increases beyond the experimental
observation time, and various concomittant  signatures of
metastability have been observed.  The resulting amorphous solids, even
though their life-time may be limited by aging and
crystallisation processes, nevertheless were found to possess a well
defined average arrested structure. 
On the theoretical side, predictions for the
glassy structure and the slow-down of the structural relaxation have
been obtained within the mode coupling theory (MCT), which describes
an idealized glass transition scenario with a divergent structural
relaxation time at a critical concentration (or temperature)
\cite{Bengtzelius84,Goetze91b,gs}. Comparisons of theory and
experimental data have shown agreement on the 30\% relative error
level  \cite{Megen91,Megen94,Beck99,Bartsch02,Goetze99}.

Considering that arrested systems like collidal glasses often are
redispersed by shaking or stirring the sample, the influence of
external shear strain or stress on glasses obviously  is of interest. 
Because glass formation is connected to a growing internal relaxation
time, an important aspect of the  imposition of external driving is
the introduction of  new time scales. In the case of
steady shearing, this aspect is generally discussed as interplay of
shear-induced and Brownian motion. Because, without shear, the glassy system
fails to reach equilibrium for long times, it is unclear to which stationary non-equilibrium state shear motion leads,
and how this non-equilibrium state approaches the equilibrium one in
the limit of vanishing shear rate. Various phenomenological models
(``constitutive equations'') either describe a yield-stress discontinuity
(Bingham plastic, Hershel-Bulkeley law), or a power-law approach (power-law fluid) \cite{larson}, while recent glass theories predict the former
\cite{Fuchs02}, the latter \cite{Berthier00}, or a transition from one to the other with concentration or temperature \cite{Fielding00}. 

In order to gain more insight into the yielding of colloidal glasses
and the nonlinear rheology of colloidal fluids, we extend here the analysis
of the microscopic approach which we recently presented
\cite{Fuchs02}. We derive from it simplified models, aiming to bring out the
qualitative and universal features.
A simple case of non-linear rheology is studied: a system of
Brownian particles in a {\em prescribed} steady shear solvent flow with constant
velocity gradient. While hydrodynamic interactions and fluctuations in the velocity profile are thus neglected from the
outset, this (microscopic) model  has the advantage that the equation of
motion for the temporal evolution of the many-particle distribution function 
can be written down exactly \cite{russel,dhont}. Our theoretical
development can therefore be crucially tested by Brownian dynamics
simulations, like Ref. \cite{Strating99},
 and constitutes a first microscopic approach
to real glassy colloidal suspensions. The properties of this
microscopic model have been worked out for low densities
\cite{Blawzdziewicz93,Bergenholtz01c} and it provides the starting
point for various (approximate) theories for intermediate
concentrations\cite{Lionberger00}.

\section{Microscopic approach}
\label{theorie}

The effect of a constant uniform shear rate $\gd$ on the particle
dynamics is measured by the Peclet number 
\cite{russel}, Pe$_0=\gd d^2/D_0$, formed with the bare diffusion
coefficient $D_0$ of a particle of diameter $d$. 
If the quiescent systems exhibits a much longer time scale $\tau$, as
do dense colloidal suspensions where $\tau$ is the final or
structural relaxation time, then a second, ``dressed'' Peclet (or
Weissenberg) number, Pe $=\gd\tau$, can be defined. 
This characterizes the influence of shear on the structural
relaxation and increases without bound at the glass transition, even
while Pe$_0 \ll 1$. In Ref. \cite{Fuchs02}, we argued that the
competition of structural 
rearrangement and shearing that arises when Pe $ > 1 \gg $ Pe$_0$ dominates the
non-linear rheology of colloids near the glass transition.

While the time scale ratios Pe$_0$ and Pe appear quite generally, the physical
mechanisms active when Pe $>1$ may be quite different for different
classes of soft matter. As we argued in the introduction, flow-induced
ordering shall here be assumed absent. In the quiescent dispersion,
the structural relaxation is dominated by (potential) particle
interactions which either cage \cite{Bengtzelius84} or bond
\cite{Bergenholtz99} a central particle among its neighbours. 
Either mechanism leads to a slowing down of particle rearrangements
accompanied by growing memory effects. The former process 
(``cage-effect'') is driven by the local order as measured in the height of the
principal peak of the static structure factor, $S_{q_p}$, and leads to a
prolonged decay-time of especially this density mode; note that its
wavevector $q_p$ is inversely related to the average particle
spacing.  The decay time of this dominant cage mode  sets the
structural relaxation time $\tau$, which in the following shall be defined by
$\Phi_{q_p}(t/\tau)=0.1$, where $\Phi_q(t)$ is the normalized
intermediate scattering function.  

It is important to realize that 
during the time interval so defined a single particle has diffused (in
direction $x$, say) a  fraction 
of its size only\cite{Megen94,Megen98,Fuchs98}: 
$\langle \Delta x^2(\tau)\rangle \approx 0.1 d^2$. In the sheared
system at Pe $=1$, therefore, it is not true that kinetic flow of
the particle with the solvent (``Taylor dispersion'' which would give
$\langle \Delta x^2(t)\rangle = \frac 23 D_0 \gd^2 t^3$ along the flow
direction \cite{dhont}) has displaced the particles relative to each
other during the time interval
$t=\tau$. This would imply a rapid destruction of the
cage already for Pe $\le1$. Rather, the effect of shear on the
structural relaxation has to be sought in its effect on the cage
mode, i.e., the collective density fluctuations with wavevector $q_p$. 
Because the steady state structure factor differs only
smoothly\cite{Bender96,Berthier02,Laun92,dhont} from the quiescent $S_q$
around $q\approx q_p$ for Pe$_0\ll1$, the effect of shear cannot lie
in a destruction of the steady state local order. This would
require larger Peclet numbers and contradict the notion that an
infinitesimal steady shear rate melts a glass. Rather, the effect
needs to arise from a shear-induced decorrelation of the memory built
up in the collective cage mode. The likely mechanism thus
appears to be the shear advection of fluctuations
\cite{dhont,Onuki79}.            
\begin{centering}
\begin{figure}[h]
\centerline{\psfig{file=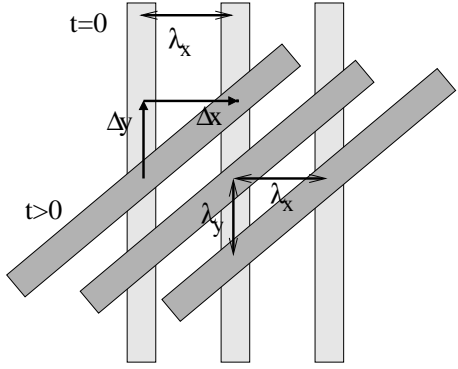,width=6.cm}}  
\caption{Advection by steady shear of a
fluctuation in $x$-direction  with wavelength $\lambda_x$ at $t=0$.
At later time $t$, its wavelength $\lambda_y$ in $y$-direction obeys:
$\lambda_x/\lambda_y=\Delta x / \Delta y= \gd t$.} 
\label{fig-1}
\end{figure}
\end{centering}

Figure \ref{fig-1} sketches, neglecting Brownian motion,
the advection of a fluctuation with
initial wavevector $(2\pi/\lambda_x,0,0)$, where $x$ points along the
flow direction, into one with wavevector
$(2\pi/\lambda_x,2\pi/\lambda_y(t),0)$, where 
$\lambda_y(t)=\lambda_x/(\gd t)$ at later time $t$. Clearly, for any
initial fluctuation with $\lambda_x\ne0$, the wavelength in the
gradient direction,  $\lambda_y(t)$, will decrease for large times.
Brownian particle motions (assisted by the interaction forces) ``smear
out'' the fluctuation with time and cause the decay of the corresponding
correlator. Because the advected wavelength decreases upon shearing,
smaller and smaller motions can cause the  fluctuation to decay.
We presume that the shear advection of density fluctuations with
wavevector $q_p$ leads to a competition between the non-linear
feedback mechanism of the cage effect and the shear-induced
decorrelation, and that this competition determines the nonlinear
rheology of concentrated particle dispersions.
Obviously, this competition involves a cooperative rearrangement of
the (finite number of correlated) particles forming the distorted
cage. Any microscopic  theory like ours therefore requires severe
uncontrolled approximations because no conceptually well-controlled
approximation scheme appears suited to this problem. 
Thus, we build upon MCT (rather than, e.g., 
approaches starting with uncorrelated binary interactions, or based on
coarse-graining procedures).

\subsection{Structural dynamics under shear}

We consider a suspension of $N$ Brownian particles, with 
density $n=N/V$, described by the Smoluchowski equation without
hydrodynamic interactions. From time $t=0$ on, 
a flow is imposed in the solvent, which points in the 
$x$- and  increases linearly along the $y$-direction:
 ${\bf v}({\bf r},t)= \kap\, {\bf
r}\, \theta(t) = \gd\, y\, \hat{e}_x\,\theta(t)$ (where $\kap$ is the
shear rate tensor, $\kappa_{ij}=\gd \delta_{ix}\delta_{jy}$,
 and $\theta(t)$ the  step function). Note that we neglect
deviations from the imposed linear flow profile and thus cannot capture
various shear-banding and other layering phenomena. (The latter may or may not arise in real experiments where only the stress, and not the velocity gradient, can be assumed constant across a sheared planar sample in steady state.) For this
situation, the Smoluchowski equation 
for the particle distribution function 
is easily formulated  \cite{Pusey85,russel,dhont}.
But, with shear, its  stationary solution 
is not known in general (except for some results at vanishing particle
concentration \cite{Blawzdziewicz93,Bergenholtz01c}), and thus steady state
quantities and correlation functions are not available. 

Recently, 
Lionberger and Russel have made progress in the semidilute
concentration regime by transferring liquid state approaches to the
steady state situation; see Ref. \cite{Lionberger00} and works cited there. 
Yet, close to the glass transition where the quiescent system develops
divergent time scales, a liquid state approach does not capture the falling out of equilibrium of the system, and 
thus (presumably) cannot handle the transition to dynamic yielding of 
a metastable solid. 
In order to capture the inherent long time scales 
we instead have suggested \cite{Fuchs02} to monitor the transient fluctuations of the
suspension after turning on the  solvent flow field at time $t=0$.
Yet this problem too cannot be solved exactly, and requires 
approximations. First, the relevant variables whose transient dynamics
shall be monitored, need to be chosen. Then equations of motion
for these variables need to be formulated. To make progress we
build upon the insights into the quiescent system provided by
the MCT  and generalize it to the out-of-equilibrium situation.

Colloidal suspensions upon densification exhibit a slowing
down of the structural relaxation (particle rearrangements) and, to
describe it, density fluctuations with wavevector $\vek q$ enter as
natural variables, $\rho_\vek{q}=\sum_{i=1}^N  
e^{i\vek{q}\vek{r}_i}$. Their magnitude is measured by the  
equilibrium structure factor $S_q=\langle \rho^*_\vek{q}
\rho_\vek{q}\rangle/N$, which changes rather little, while their dynamics
slows down dramatically.
We follow MCT in  considering the set of
density fluctuations as a set of slow variables. Thus, arbitrary steady-state expectation values are obtained from determining the overlap of
the relevant quantities with density fluctuations.
This requires us to find the transient density fluctuations. They shall be 
 determined from a  closed set of equations of motion for the  
intermediate scattering functions, which is obtained by performing
a mode coupling approximation.
 
With these approximations, various steady state quantities, like
 the  thermodynamic shear stress \cite{Batchelor77} $\sigma=\langle
\sigma_{xy}/V \rangle^{(\gd)}$   can be  calculated. (Note that steady state averages are abbreviated by $\langle\ldots
\rangle^{(\gd)}$, while equilibrium ones without shear are given by
 $\langle\ldots \rangle$.) The transverse stress also provides the shear 
viscosity which follows as $\eta(\gd) = \eta_\infty+\sigma/\gd$, where
the solvent viscosity is denoted $\eta_\infty$. Our final microscopic
expression for the steady state stress 
$\sigma$ is found to be:
\beq{mi1}
\sigma = \langle \sigma_{xy}/V \rangle^{(\gd)}  \approx 
\frac{k_BT \,\gd}{2}  \int_0^\infty\!\!\!\!dt\, 
\int\!\!\frac{d^3k}{(2\pi)^3}\;
\frac{k_x^2k_yk_y(t)}{k\, k(t)}\;
\frac{S'_kS'_{k(t)}}{S^2_{k(t)}}\; \Phi^2_{\vek{k}}(t)\; ,
\eeq
with $t$ the time since switch-on, $k_BT$ the thermal energy,
 and $S'_k=\partial S_k/\partial k$. 
The transient density fluctuations are given by 
$\Phi_{\vek{k}}(t)=
\langle \rho_{\vek{k}(t)}^*(t) \rho_\vek{k}
 \rangle/(NS_k)$ and are 
normalized by the quiescent $S_k$. Because of shear advection,  time
dependent wavevectors 
$k(t)=\sqrt{k^2+2k_xk_y\gd
t+k_x^2\gd^2t^2}$ appear; a 
fluctuation with wavevector $\vek k$ at the start of shearing $t=0$
has a finite overlap with a fluctuation 
at the advected wavevector $\vek{k}(t)=
\vek{k}+\vek{k}\kap \,t$ at the later time $t>0$.

In summary,
steady state quantities shall be determined by considering the
structural relaxation under shearing and ``integrating through the 
transient dynamics''.
Because Pe$_0\ll1$, we expect ordering or layering transitions to be
absent \cite{Laun92}; and as hydrodynamic interactions are presumed to
play a subordinate role during the structural relaxation
\cite{Bergenholtz01b} we neglect these too, focusing solely on the
Brownian contribution to the transverse (shear) stress. 
A major approximation within our approach entails the
elimination of explicit particle forces  in favour of the
quiescent-state structure factor $S_q$ (the only input in our theory
and taken to be known). This is a
near-equilibrium assumption  that is formally uncontrolled
but motivated, at least in part, by the smallness of Pe$_0$.   
Because we approximate nonlinear couplings under shear using 
equilibrium averages, we require the system to remain ``close to
equilibrium'' in some sense. We leave to future work any attempt to make this sense more precise and, if possible, to relax the assumption involved.

\subsection{Glass stability analysis}

Little can be gained in our approach without specifying the
dynamics of the transient intermediate scattering 
functions which describe how the equilibrium structure 
changes with time into that of a steadily sheared state.
Obviously, any uncontrolled approximation (like mode coupling), which we are
now forced to perform to obtain the  
equations of motion for the
transient fluctuations, can introduce errors
of unknown quality and magnitude into our 
results. But while the quantitative accuracy of the equations we propose in Ref.\cite{Fuchs02}
has not been tested yet, there are qualitative 
conclusions which can be drawn from the structure of the equations,
which are rather independent of the  
microscopic details of the approximation. Thus, in order to test our
basic approach, these more universal  
aspects are of central interest and should be the ones chosen for
initial comparison to experiments or 
simulations. We will consider only these aspects here, but in later
sections two simple 
models are used to study how far the universal 
aspects dominate the model-dependent results.

The universal results  which follow from our approach are connected to
the stability equation of a (quiescently) 
arrested glassy structure which gets melted away by the imposed shear.
From an expansion of the transient  
density fluctuations around the initially arrested structure close to the glass
transition we get the bifurcation 
equations which describe how the localization driven by the cage
effect competes with the fluidization induced by 
shearing. The derivation of a generalized  ``factorization theorem"
and the so-called $\beta$-scaling equation, proceeds  
by a straightforward perturbation calculation which is given in Appendix
A. Close to the  
bifurcation, the transient density fluctuations are given by one
function ${\cal G}(t)$, which depends on control parameters,
and determines the dynamics on all length scales:
\beq{mi2}
\Phi_q(t) = f^c_q + h_q \,   {\cal G}(t)\; .
\eeq
The numbers $f_q^c$ describe the glassy structure at the instability and
the critical amplitude $h_q$ is connected to the cage-breaking
particle motion; both retain their definition from the unsheared situation.  
The function ${\cal G}(t)$ contains the essential
non-linearities of the bifurcation dynamics which arise from the physical
feedback mechanism (the cage effect) and the shear disturbance. It
depends on a few material parameters only, and follows from
\beq{mi3}
\varepsilon - c^{(\gd)} \; (\gd\; t)^2 + \lambda\; {\cal G}^2(t) =
\frac{d}{dt} \; \int_0^tdt'\; {\cal G}(t-t')\; {\cal G}(t') \; ,
\eeq
where $\varepsilon,\lambda$ and $c^{(\gd)}$ are defined in Appendix A. The so-called separation
parameter $\varepsilon$ measures the distance to the transition 
and the exponent parameter  $\lambda$ determines power-law exponents
resulting from \gl{mi3}, and is known for some systems \cite{Bengtzelius84}.
The shear-related  parameter $c^{(\gd)}$ is a (relatively unimportant) number
of order unity, which sets the scale for $\gd$ and could be absorbed
into an effective shear rate $\sqrt{c^{(\gd)}}\,\gd$. 

The two derived equations, \gls{mi2}{mi3},
describe an expansion around the transition point between 
a non-Newtonian fluid and a yielding solid (within our approach, this applies whatever microscopic model is chosen) 
where divergent relaxation times arise from self-consistently
calculated memory effects and compete with an externally imposed
(shear-driven) loss of memory.
Corrections of higher order in the small quantities ($\varepsilon,
\gd, {\cal G}$) are  neglected in \gls{mi2}{mi3}; 
see Ref.\cite{Franosch97} for  a calculation of the leading corrections
and for background on \gl{mi3} at $\gd=0$.
In the following sections we will discuss two simplified models
which, on the one hand, allow a more detailed analysis, and on the
other hand share the universal stability properties derived from
 \gls{mi2}{mi3}.

\section{Models}

We now present two, progressively more simplified,  models that provide insights into the
generic scenario of non-Newtonian flow, shear melting and solid yielding
which emerge from our approach.

\subsection{Isotropically sheared hard sphere model (ISHSM)}

On the fully microscopic level of description of a sheared
 colloidal suspension, kinetic
flow of the particles with the solvent leads to anisotropic dynamics.
Yet, because of the strong hindering of the motion
at high densities, which leads to a caging of particles, the 
development of the anisotropic ``Taylor dispersion'' may not yet be
important at Pe $\simeq 1$ so that the motion stays locally isotropic. Recent simulation data of
steady state structure factors support this consideration and
indicate  an isotropic distortion
of the structure for Pe$_0\ll1$, while the Weissenberg number Pe
is already large 1 \cite{Berthier02}. A mechanism 
which operates independently from Taylor dispersion
arises from the  shift of the advected wavevectors with time to higher values.
As the effective potentials felt by density fluctuations
evolve with increasing wavevector, this leads to 
a decrease of friction functions, speed-up of structural
rearrangements and shear-fluidization.
Therefore, one may hope that a model of isotropically sheared hard
spheres (ISHSM), which  for
$\gd=0$  exhibits the nonlinear coupling of density
correlators with wavelength equal to the average particle distance
(viz. the ``cage-effect''),
and which, for $\gd\ne0$, incorporates shear-advection, may be not too
unrealistic.

Thus, in the ISHSM, the equation of motion for the density fluctuations at time $t$
after starting the shear is approximated by the  one of the
quiescent system:
\beq{iso1}
\dot{\Phi}_q(t) + \Gamma_q  \left\{
\Phi_q(t) + 
 \int_0^t\!\!\! dt'\; m_q(t-t') \,
\dot{\Phi}_q(t')
\right\} = 0 \; .
\eeq
where $\dot{\Phi}_q(t)=\partial_t \Phi_q(t)$, and
the initial decay rate is $\Gamma_q=q^2 D_0 / S_q$ with the single
particle Brownian diffusion coefficient given by the solvent
viscosity, $D_0=k_BT/(3\pi\eta_\infty d)$.
The memory function also is taken from the unsheared situation but
now shear advection is entered into the overlap of the fluctuating
forces with density fluctuations at time $t$:
\beq{iso2}
 m_q(t) \approx 
\frac{1}{2N} \sum_{\vek{k}}  V^{(\gd)}_{q,\vek{k}}(t)\;
 \Phi_k(t) \, \Phi_{|\vek{q}-\vek{k}|}(t) \; ,
\eeq
with
\beq{iso3}
V^{(\gd)}_{q,\vek{k}}(t) = \frac{n^2S_qS_kS_p}{q^4}
\left[ \vek{q}\cdot\vek{k}\; c_{k(t)} + \vek{q}\cdot\vek{p}\; c_{p(t)}  \right]
\left[ \vek{q}\cdot\vek{k}\; c_{k} + \vek{q}\cdot\vek{p}\; c_{p} \right]
\eeq
where $\vek{p}=\vek{q}\!-\!\vek{k}$, and the length of the advected
wavevector is approximated by
$k(t)=k(1+(t\gd)^2/3)^{1/2}$. The effective potentials are given
by the direct correlation functions, $c_q=(1-1/S_q)/n$.
(For background on this model without shear, see Refs. 
\cite{Bengtzelius84,Goetze91b,Franosch97}.) 

For hard spheres, the quiescent $S_q$, taken from Percus-Yevick
theory\cite{russel},  depends only on the packing
fraction $\phi$; discretising the wavevector
integrations in \gl{iso2} as done in Ref.\cite{Franosch97}, we find that the model's glass
transition lies at $\phi_c=0.51591$. Thus $\eps$ (where
$\eps\simeq 3.0(\phi-\phi_c)$),
and $\gd$ are the only two control parameters determining the rheology.
The exponent parameter becomes $\lambda=0.74$ and
$c^{(\gd)}\approx$ 3, while $S_q$ has its peak at $q_p=7/d$.

In the same spirit of incorporating advection as the only effect of
shearing, the expression  for the transverse modulus may be simplified to
\beq{iso4}
\sigma=\frac{k_BT\, \gd}{60\pi^2}\;
\int_0^\infty\!\! dt\; \int\!\! dk\; k^4
  \frac{S'_kS'_{k(t)}}{S^2_{k(t)}}\; \Phi_k^2(t)  \; .
\eeq
The resulting numerical results for the ISHSM are discussed below (section IV).

\subsection{Schematic $F_{12}^{(\gd)}$ model}

The central features of the equations of motion of the ISHSM are that
it reproduces the stability equation from the microscopic approach,
\gl{mi3}, and that the vertex $V^{(\gd)}$ contains the competition of
two effects. First, it increases with increasing particle
interactions (``collisions'' or ``cage effect'') which leads to a
non-ergodicity transition in the absence of shear, and second, it
vanishes with time because of shear-induced decorrelation. Both these
effects can be captured in an even simpler ``schematic''
model, which moreover can be made to obey \gl{mi3} also. This schematic
F$_{12}^{(\gd)}$ model considers one normalized correlator $\Phi(t)$,
with $\Phi(t\to0)=1-\Gamma t\ldots$,
 which obeys a generalized relaxation equation:
\beq{sm1}
\dot{\Phi}(t) + \Gamma  \left\{
\Phi(t) +  \int_0^t\!\!\! dt'\; m(t-t') \, \dot{\Phi}(t')
\right\} = 0 \; .
\eeq
Again, in the absence of the memory kernel $m$ the dynamics is trivial,
$\Phi(t)=\exp({-\Gamma t})$, and a low order polynomial ansatz for $m$
suffices to  study the generic schematic model. We choose
\beq{sm2}
 m(t) = \frac{1}{1+(\gd t)^2} \; 
 \left(v_1 \Phi(t)+v_2 \Phi^2(t) \right) \; .
\eeq
Without shear, this model has been studied extensively
\cite{Goetze84,Goetze91b}. Increasing particle caging is modeled by
increasing coupling parameters $v_1,v_2\ge0$, and the only effect of
shearing is to cause a time dependent decay of the friction kernel $m$.
The system loses memory because of shearing.
 The role of the transport coefficient
(viscosity) $\eta$ is played by the average relaxation time obtained
from integrating the correlator, and this also is taken to determine the stress:
\beq{sm3}
\sigma = \gd\;  \eta = 
\gd \; \langle \tau \rangle = \gd \int_0^\infty\!\!\! dt\;  \Phi(t) \; .
\eeq
For the parameters of the model, values studied in the literature shall
be taken.
While the parameter $\Gamma$ just sets the time scale, the 
two interaction parameters are chosen as $v_2=v_2^c=2$ and
$v_1=v_1^c+\delta v_1$, where $v_1^c=v_2^c(\sqrt{4/v_2^c}-1)\approx
0.828$. From an analysis similar to Appendix A, the critical
non-ergodicity parameter is found as
$f_c=1-\frac{1}{\sqrt{v_2^c}}\approx0.293$ and the parameters in
\gl{mi3} emerge as follows:  
\beq{sm4}
\lambda=1-f_c\approx0.707\;,
\quad \eps=\frac{\delta v_1 f_c+\delta v_2 f_c^2}{1-f_c}=\frac{\delta
v_1 f_c}{1-f_c}\; ,\quad 
c^{(\gd)} = \frac{v^c_1 f_c+v^c_2 f_c^2}{1-f_c}\approx 0.586\; .
\eeq
The major advantage of this schematic model, besides its numerical
simplicity, is that it encodes the two physical mechanisms at work in
two handy parameters. The one, $\eps$, parametrises the
tendency of the undriven system to arrest, while the other, $\gd/\Gamma$,
measures the loss of memory brought about by shearing.

\section{Results and discussions}

Both models described above exhibit a non-ergodicity transition which corresponds to
an ideal fluid-to-glass transition within MCT.
 For $\gd=0$, upon smooth changes of the input equilibrium
state parameters, a fluid 
with $\Phi_q(t\to\infty)\to0$ turns into an amorphous solid,
$\Phi_q(t\to\infty)\to f_q>0$. The $f_q$ are called glass form factors
and describe the arrested structure. While transport
coefficients of the fluid, like the viscosity, are connected to the
longest relaxation  time $\tau$ of the $\Phi_q(t)$, elastic constants of the
solid, like the  transverse elastic modulus $G_\infty$, are given by the
$f_q$. (For some in-depth, albeit somewhat older, discussions of the idealized MCT see
Refs. \cite{Goetze91b,gs}, and for a more recent discussion of
experimental tests Ref. \cite{Goetze99}.)
\begin{centering}
\begin{figure}[h]
\centerline{\psfig{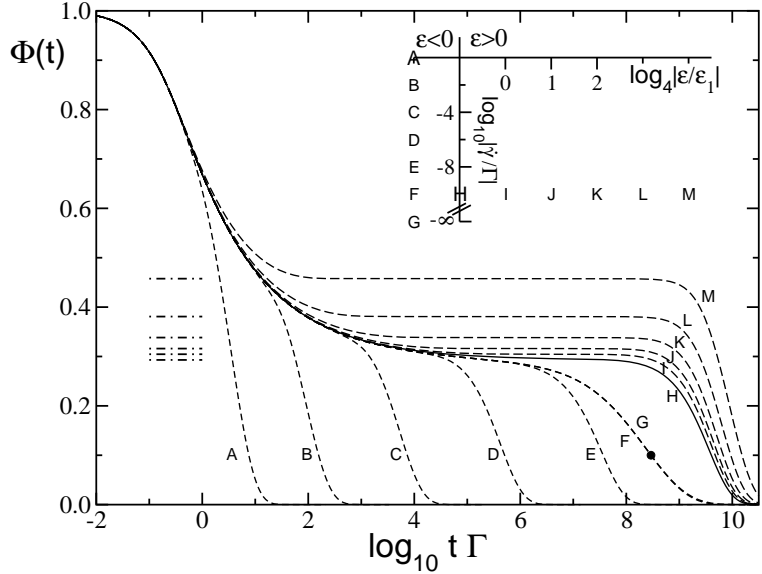}}  
\caption{Correlators $\Phi(t)$ of the schematic $F_{12}^{(\gd)}$ model
for the
separation parameters $\varepsilon$ and shear rates $\gd/\Gamma$ as
marked in the inset; curves A to F for
$\eps=-\eps_1=-10^{-3.79}$ and $\gd/\Gamma=10^{-n}$ with
$n=0,2,\ldots,10$, while curve G corresponds to an
unsheared fluid $\gd=0$ at $\varepsilon=-\varepsilon_1$ (a filled
circle marks where the final relaxation time $\tau=10^{8.5}/\Gamma$
is read off, $\Phi(t=\tau)=0.1$). Curve  H lies
at the critical point $\varepsilon=0$ for $\gd/\Gamma=10^{-10}$, while
I to M keep $\gd/\Gamma=10^{-10}$ but increase $\eps=4^n\eps_1$,
$n=0,\ldots,4$. 
The horizontal bars indicate the heights of the non-ergodicity
parameters $f$  for $\eps\ge0$ (compare curves H to
M),  which would be approached at long times for $\gd=0$,
$\Phi(t\to\infty)=f$.} 
\label{fig0}
\end{figure}
\end{centering}

\subsection{Transient fluctuations}
\label{transdens}

Figure \ref{fig0} shows correlators of the $F_{12}^{(\gd)}$ model for
various shear-rates and distances from the non-ergodicity transition at
$\eps=0$. Curve G there corresponds to a non-sheared fluid state close
to the transition, $\eps=-\eps_1=-10^{-3.79}$ and $\gd=0$. This shows the typical
two-step relaxation pattern with  microscopic short-time dynamics for
$t\Gamma={\cal O}(1)$, followed by the approach to an intermediate
plateau at $f_c\approx 0.293$, and the final decay characterised by the
(final or $\alpha$-) relaxation time $\tau$. For positive
separation parameters, $\eps\ge0$, the final decay is absent and the
correlators   approach finite long time  limits $\Phi(t\to\infty)=f\ge
f_c$ (not shown; these $f$'s are indicated by horizontal bars at the left of the figure).
Including a finite shear rate corresponding to a small but finite bare
Peclet number Pe$_0=\gd/\Gamma$ in the model, little effect of shear
on the short time dynamics is seen because of Pe$_0\ll1$. A drastic effect on
the final decay, however, is seen in curves A to E, because the dressed
Peclet (or Weissenberg) number Pe$=\gd\tau$ is not negligible.  Moreover, all
glassy curves ($\eps\ge0$), which would stay arrested for $\gd=0$,
are seen to 
decay by a shear-induced process whose time scale is set by the
inverse shear rate, and whose amplitude depends on the distance to the
transition.
\begin{centering}
\begin{figure}[h]
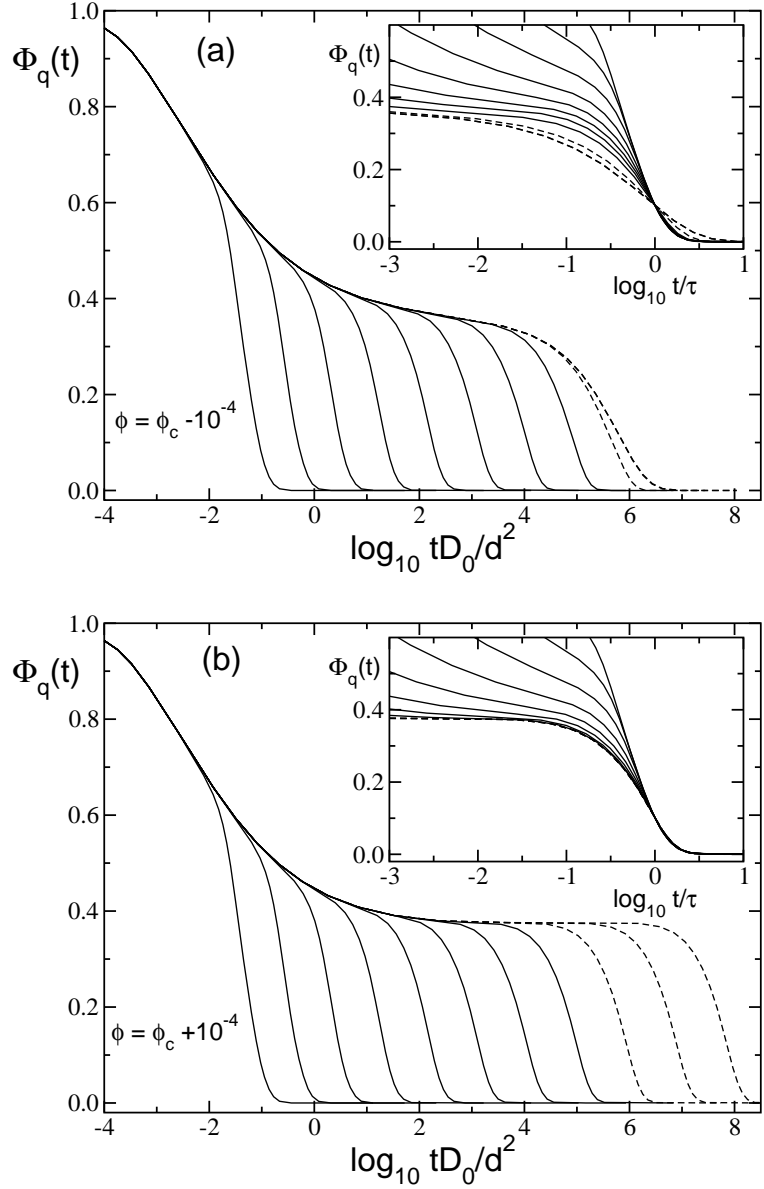

\centerline{\psfig{file=ishsmphivontfluidk.eps,width=10cm}} 
\vspace*{0.5cm}
\centerline{\psfig{file=ishsmphivontglassk.eps,width=10cm}}  
\caption{Normalized transient density correlators $\Phi_q(t)$ of the
ISHSM at wavevector $q=3.4/d$ below (panel (a) at
$\phi=\phi_c-10^{-4}$) and above  (panel (b) at
$\phi=\phi_c+10^{-4}$) the transition for increasing shear rates
Pe$_0=9^n*10^{-8}$ with $n=0,\ldots,10$
from right to left; the distances correspond to $\eps=\pm
10^{-3.53}$. Curves for $n=$ 9, 10 carry short and for $n=$ 8 long
dashes; note the collapse of the two short dashed 
curves in (a). The insets show the data rescaled so as to coincide
at $\Phi(t=\tau)=0.1$. }
\label{fig1}
\end{figure}
\end{centering}

The stability analysis of \gls{mi2}{mi3} describes the correlators for
a window around $\Phi_q\approx f_q^c$ and a finite window in time,
which both can be estimated from the condition $|{\cal G}(t)|\ll1$; 
for $\gd=0$ these have been worked out in detail \cite{Franosch97}.
As the analysis in Appendix B shows,
  $(\gd t)^2$ dominates for long times in \gl{mi3}, and therefore always
\beq{re1}
{\cal G}(t\to\infty)\to-t/\tau^{(\gd)}\; ,\quad\mbox{with }\quad
\tau^{(\gd)}=\sqrt{\frac{\lambda-1/2}{c^{(\gd)}}} \;  \frac{1}{|\gd|} \;.
\eeq
Hence, under the imposed shear
flow, density fluctuations always decay, as this decrease of ${\cal G}(t)$ for
long times initiates the final relaxation of
$\Phi_{\vek{q}}(t)$ to zero. (In this region the corrections to
\gl{mi2} of higher order in $\cal G$  become important.) 
Arbitrarily small steady shear rates $\gd$ melt the glass, as expected, and
for small Pe$_0$ there appears a separation of time scales between
short-time motion and the shear induced final decay; see
curves H  to M in figure \ref{fig0}.

For more detailed insight into the shapes of the relaxation curves,
we turn to the ISHSM.
Figure \ref{fig1} displays density correlators at two densities,
just below (panel (a)) and just above (panel (b)) the
transition,  for varying
shear rates. Again, in almost all cases Pe$_0$ is negligibly small and
the short-time dynamics is not affected. In the fluid case, the
final or $\alpha$-relaxation is also not affected for the two smallest Pe
values, but for larger Pe the non-exponential $\alpha$-relaxation becomes
faster and less stretched; see the inset of fig. \ref{fig1}(a). 
(While the techniques of Appendix B allow us to discuss 
this shape change, we do not do so here for lack of space.)
The glassy curves at $\eps>0$, panel (b), exhibit a  shift of the
final relaxation with $\tau^{(\gd)}$ and asymptotically approach a
scaling function $\Phi_q^+(t/\tau^{(\gd)})$.
The master equation for the ``yielding''
scaling functions $\Phi_q^+$ in the ISHSM can be
obtained from eliminating the short-time dynamics in \gl{iso1}. After
a partial integration, the equation with $\dot\Phi_q(t) = 0$ is solved by the scaling functions:
\beq{re2}
\Phi_q^+(\tilde t) = m^+_q(\tilde t) - \frac{d}{d\tilde t}\,
\int_0^{\tilde t}d\tilde t'\; m^+_q(\tilde t-\tilde t')
\Phi_q^+(\tilde t')\; , 
\eeq
where $\tilde t=t/\tau^{(\gd)}$, and the memory kernel is given by
\beq{re3}
 m^+_q(\tilde t) = 
\frac{1}{2N} \sum_{\vek{k}}  
V^{(\tilde{\gd})}_{q,\vek{k}}(\tilde t)\;
 \Phi^+_k(\tilde t) \, \Phi^+_{|\vek{q}-\vek{k}|}(\tilde t) \; .
\eeq
While the vertex is evaluated at fixed shear rate, 
$\tilde{\gd}=\sqrt{\frac{\lambda-\frac 12}{c^{(\gd)}}}$, it depends on
the equilibrium parameters.  The corresponding results for the 
$F_{12}^{(\gd)}$ can easily be obtained. 

The importance of the $\beta$-correlator $\cal G$ for the derivation
of the above 
scaling functions is because $\cal G$, according to \gl{re1}, depends on
$\tau^{(\gd)}$ and not on internal time scales like $\tau$.
It provides the initial conditions for \gl{re2}, which follow from the
analysis of 
\gl{mi3} in 
Appendix B. In the glass, $\eps>0$, \gl{b5} leads to 
\beq{re4}
\Phi_q^+(\tilde t\to0) = f^c_q + h_q \; \sqrt{\frac{\eps}{1-\lambda}}\;
\left[ 1 - \frac{1-2\lambda}{4\eps} \; \tilde t^2 + {\cal
O}(\frac{1}{\eps^2}(\tilde t)^4) \right] \;,
\eeq 
and for somewhat longer times, $\tilde t \gg \sqrt\eps$,  according to
\gl{b9}, this merges into:
\beq{re5}
\Phi_q^+(\sqrt\eps\ll \tilde t\ll1) = f^c_q - h_q \; \tilde t \;.
\eeq
At the transition, $\eps=0$, \gl{b8} shows that \gl{re5} holds down to
$\tilde t\to0$. The scaling functions defined thereby describe the yielding
behavior of a sheared glass where the relaxation time is set by the
external shear rate and internal relaxation mechanisms are not active.
\begin{centering}
\begin{figure}[h]
\centerline{\psfig{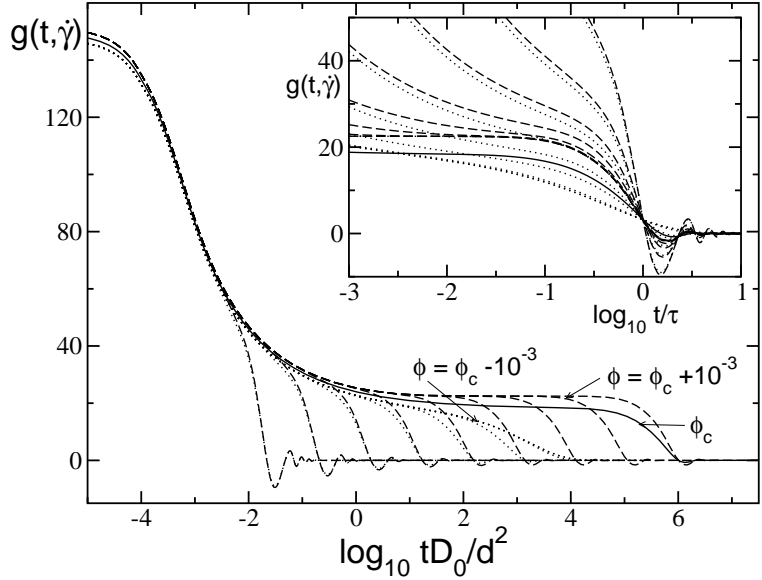}} 
\caption{Transient non-Newtonian shear modulus $g(t,\gd)$ of the ISHSM
in units of  $k_BT/d^3$ for the packing fractions
 $\phi=\phi_c\pm10^{-3}$ ($\eps=\pm
10^{-2.53}$;  dashed/ dotted lines, respectively)
 for increasing shear rates Pe$_0=9^n*10^{-6}$ with $n=0,\ldots,8$
from right to left; note the collapse of fluid lines for the smallest
Pe$_0$.  The solid line gives $g(t,\gd)$ for $\phi=\phi_c$ and  Pe$_0=10^{-6}$.
The inset shows the data rescaled so as to coincide
at $g(t=\tau,\gd)=5$; note the collapse of the $n=$ 6,7 \& 8 curves for
both $\phi>\phi_c$ and $\phi<\phi_c$. }
\label{fig2}
\end{figure}
\end{centering}

Figure \ref{fig2} shows the transient shear modulus $g(t,\gd)$ of the
ISHSM which determines the viscosity via $\eta=\int_0^\infty\!\! dt \,
g(t,\gd)$. It is the time derivative of the shear stress growth function
$\eta^+(t,\gd)$  (or transient start up viscosity; here, the $^+$
labels the shear history\cite{larson}),  $g(t,\gd)=\frac{d}{dt}
\eta^+(t,\gd)$, and in the Newtonian-regime reduces to the time
dependent shear modulus, $G(t)=g(t,\gd=0)$.
The $g(t,\gd)$ shows all the features exhibited by the correlator of
the schematic 
model, and by the density correlators in the ISHSM, and thus the
discussion based upon 
$\cal G$ and the yielding scaling law carries over to it. But in
contrast to the correlator of the $F^{(\gd)}_{12}$, and 
more so than the $\Phi_q$, the function $g(t,\gd)$ becomes
negative (oscillatory) in the final approach towards zero, an effect
more marked at high Pe. This  behavior originates in the general
expression for $g(t,\gd)$, \gl{mi1}, where the vertex reduces to a 
positive function (complete square) only in the absence of shear
advection. A overshoot and oscillatory approach of the start up
viscosity to the steady state value,
$\eta^+(t\to\infty,\gd)\to\eta(\gd)$, therefore are  generic
features predicted from our approach.

In the discussion that follows, the leading corrections to the scaling law
for the yielding correlators will be required.
The analysis of \gl{mi3} suggests that to next order for $\eps\ge0$:
\beq{re6}
\Phi_q(t) \to \Phi^+_q(\tilde t) + \delta(\eps,|\gd t_0|)\; 
\bar{\Phi}^+_q(\tilde t) + {\cal O}(\delta^{2}) \; ,
\eeq
where $t_0$ is a matching time, and the correction
$\bar{\Phi}^+_q(\tilde t)$ exhibits a weak divergence  for short rescaled 
times,
 $\bar{\Phi}^+_q(\tilde t\to0)\propto \tilde
t^{-(2\lambda-1)/\lambda}$,  as follows from \gls{b8}{b9}. The inset of fig. \ref{fig1}(b) shows the rise of the
correlators above the yielding master function at short times and for
very small Pe$_0$.  Except for
the fact that it is integrable, this function is of no further interest here. 
The small parameter $\delta$, however, sets the scale of the
corrections and (as shown in \gls{b8}{b9}) exhibits the following scaling properties 
\beq{re7}
\delta \propto \left\{\begin{array}{lcl}
|\gd t_0 |^m & \mbox{ with }\,
m=\frac{a}{1+a}\frac{2\lambda-1}{\lambda} &\mbox{for }\, \eps=0 \\
\eps^{m'} &  \mbox{ with }\, m'=\frac{2\lambda-1}{2\lambda}
&\mbox{for }\,  |\gd t_0|^{\frac{2a}{1+a}} \ll \eps \ll 1 \end{array}\right.
\eeq 
These results will be used below.

\subsection{Flow curves}
\label{consteq}

\begin{centering}
\begin{figure}[h]
\centerline{\psfig{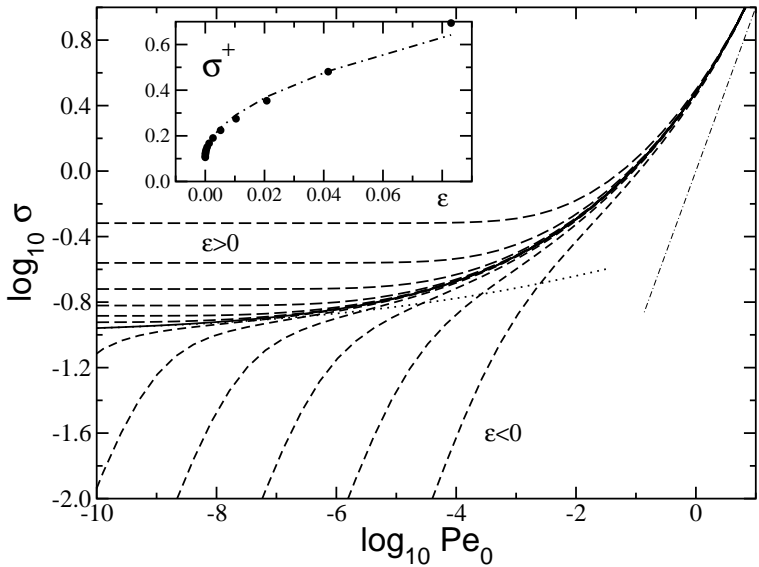}}  
\caption{Steady state ``stress'' $\sigma=\gd\langle \tau\rangle$ in the
schematic $F_{12}^{(\gd)}$ model as function of Pe$_0=\gd/\Gamma$ at
various distances from  the transition: $\eps=0$ (solid
line)  and  $\eps = \pm 4^n \eps_1$ for $n=-1,0,..,4$ (while
$\eps_1=10^{-3.79}$)  with short
(long) dashes for fluid $\varepsilon<0$ (glassy $\varepsilon>0$) curves.
At $\varepsilon=0$, a
 dotted line gives $\sigma=\sigma^+_c(1+2.48\,
\gd^{0.143})$  for $\gd\to0$, with $\sigma_c^+=0.10$,
 while a thin dot-dashed line gives $\sigma_c=\gd$ for
$\gd\to\infty$. The inset shows the finite limits
$\sigma^+=\sigma(\varepsilon\ge0,\gamma\to0+)$ obtained above the
transition; the dot-dashed curve is $\sigma^+=\sigma^+_c+1.88\sqrt{\varepsilon}$.}
\label{fig3}
\end{figure}
\end{centering}

\begin{centering}
\begin{figure}[h]
\centerline{\psfig{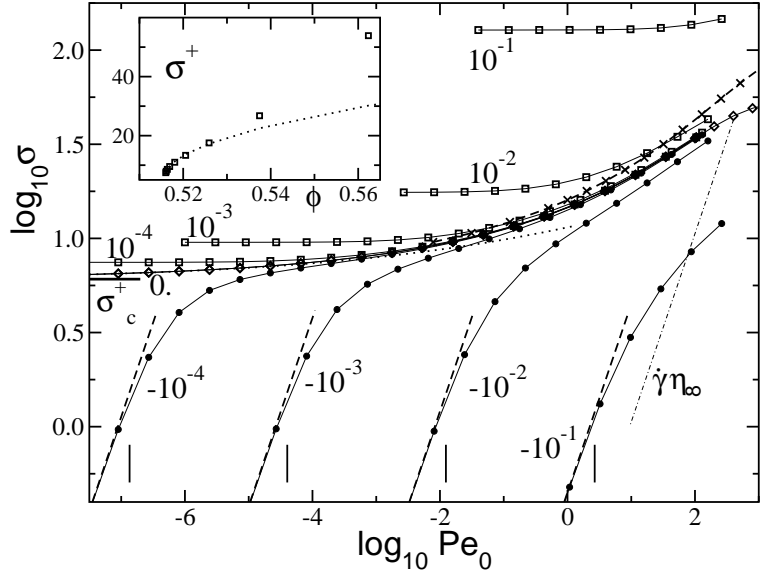}}  
\caption{Steady state shear stress
 $\sigma$  in units of $k_BT/d^3$ 
 versus Pe$_0=\gd d^2/D_0$, for the ISHSM
at various distances from its glass
transition, $\phi-\phi_c$ as labeled; circles correspond to fluid,
diamonds to the critical, and squares to glassy densities. For the fluid cases,
$\phi<\phi_c$,  
dashed lines indicate Newtonian fluid behavior, $\sigma=\eta\gd$,
while vertical bars mark Pe$=\gd\tau=1$, with the structural relaxation
time taken from $\Phi_{q=7/d}(t=\tau)=0.1$.
The stress which would additionally arise from the background solvent
viscosity, $\sigma=\gd \eta_\infty$, is marked by a dot-dashed line.   For the
critical density, $\phi_c$, the critical yield stress, $\sigma^+_c=6.04$,
is shown by a horizontal bar, and the dotted line
$\sigma=\sigma^+_c(1+0.89\, \gd^{0.152})$ holds for $\gd\to0$.
Crosses connected by a thick dashed line give $\sigma$ at
$\phi=\phi_c$ but for a different numerical discretization of the
memory kernel (3-times larger cut-off $k_{\rm max}$). The inset shows
the rise of the dynamical yield stress
$\sigma^+=\sigma(\varepsilon\ge0,\gd\to0+)$ in the glass together with
the fitted square
root asymptote, $\sigma^+=\sigma_c^++112\sqrt{\phi-\phi_c}$.}
\label{fig4}
\end{figure}
\end{centering}

\begin{centering}
\begin{figure}[h]
\centerline{\psfig{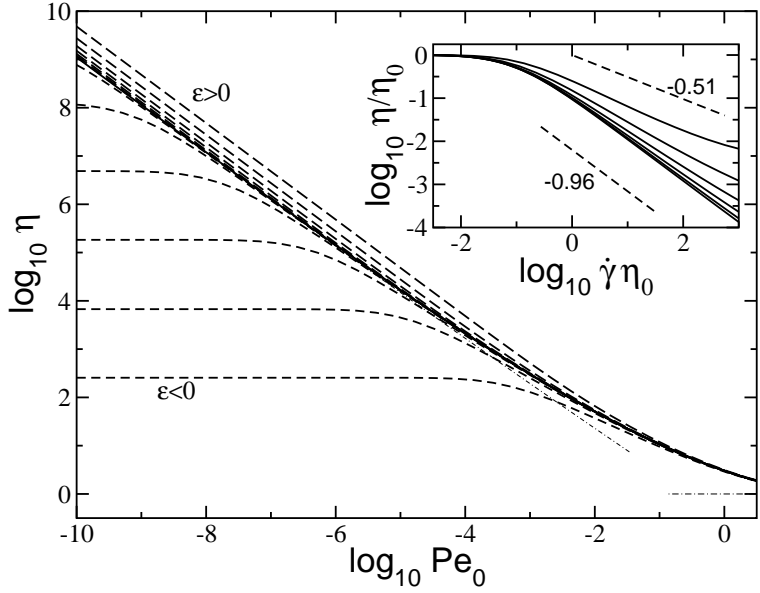}}  
\caption{Shear dependent ``viscosity'' $\eta=\langle \tau\rangle$ (in units
of $1/\Gamma$) in the schematic $F_{12}^{(\gd)}$ 
model as function of Pe$_0=\gd/\Gamma$ at various distances from
the transition (data and line styles as in figure \protect\ref{fig3}).
The inset shows the fluid curves $\varepsilon<0$ rescaled with the
viscosity $\eta_0=\eta(\gd=0)$ of the unsheared situation.  Two power-laws
(dashed lines) indicate the minimal and maximal slopes of the curves
around $\gd\eta_0=10^2$.}
\label{fig5}
\end{figure}
\end{centering}

\begin{centering}
\begin{figure}[h]
\centerline{\psfig{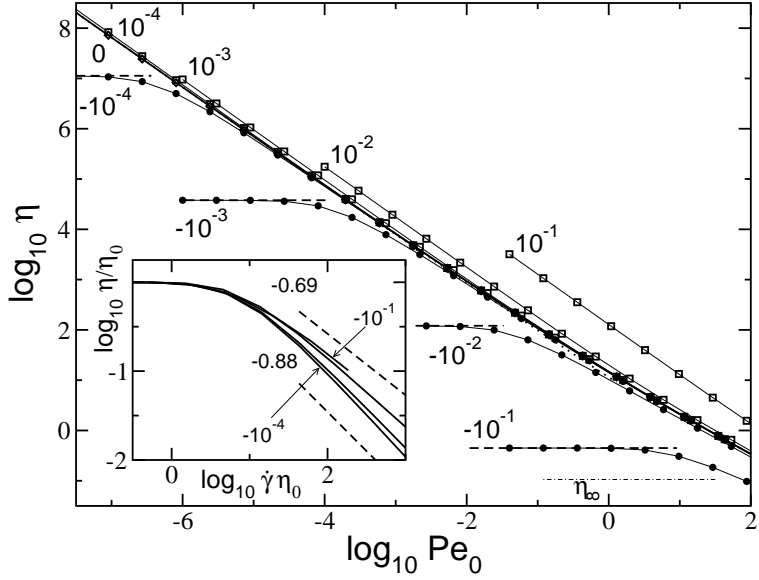}}  
\caption{Steady state non-Newtonian shear viscosity
 $\eta$  in units of $k_BT/D_0d$ 
 versus Pe$_0=\gd d^2/D_0$, for the ISHSM
at various distances from its glass
transition, $\phi-\phi_c$ as labelled; data and line styles from figure
\protect\ref{fig4} (the dot-dashed horizontal line gives the neglected
background solvent viscosity). 
The inset shows the fluid curves $\varepsilon<0$ rescaled with the
viscosity $\eta_0=\eta(\gd=0)$ of the quiescent fluid. Two power-laws
(dashed lines) indicate the minimal and maximal slopes of the curves
around $\gd\eta_0=10^2$.}
\label{fig6}
\end{figure}
\end{centering}

In our approach, steady state properties of the sheared system are
obtained via time integrals --- from switching on the rheometer at $t=0$
up to very late times when the system has relaxed into the
non-equilibrium stationary state. The evolution of the system is
approximated by following the transient density fluctuations, which were
discussed in the previous section for two specific simplified models.
Equation (\ref{mi1}) gives the thermodynamic shear stress
in our fully microscopic approach, while \gls{iso4}{sm3} give
simplified expressions for it using the two models.  Inserting the transient
density correlators from section \ref{transdens}, the stress versus
strain rate curves (``flow curves'') of the two models can be discussed. Such
relations $\sigma(\gd)$ often are postulated en route to deriving phenomenological 
``constitutive equations'' for nonlinear flow behaviour, whereas our approach leads, in principle, to their microscopic 
derivation. Figures \ref{fig3} to \ref{fig6} present the results, the
first two as plots of $\sigma$ versus $\gd$, while the second pair
shows $\eta$ versus $\gd$. We will discuss the general qualitative
features jointly for both models.

In the fluid, we find a  linear or Newtonian regime in the limit
$\gd\to0$, where we recover the standard
MCT approximations for the transport coefficients of a viscoelastic
fluid \cite{Goetze91b,gs}. Hence $\sigma \to \gd \,\eta(\gd=0)$ holds for Pe
$\ll 1$, as shown in figure \ref{fig4}, where Pe calculated with the
structural relaxation time $\tau$ is included. As discussed in
section \ref{theorie}, the growth of $\tau$ (asymptotically) dominates
all transport coefficients of the colloidal suspension and causes an
proportional increase in the viscosity $\eta$. For Pe $>1$, the
non-linear viscosity shear thins, and $\sigma$ increases sublinearly
with $\gd$. Without analysing these complicated flow curves in detail
here, we note that additional, shorter time 
scales than $\tau$ enter; these cause the shape change of the density
correlators shown in figure \ref{fig2}(a) and also affect $\sigma$
and $\eta$. The stress versus strain rate figures \ref{fig3},
\ref{fig4} clearly exhibit a broad crossover between the linear
Newtonian and a much weaker (asymptotically)  $\gd$-independent
variation of the stress. 

Replotting the identical data as
viscosity versus strain rate (figures \ref{fig5}, \ref{fig6}) 
these subtle features get compressed by the requirement of a much
larger scale range on the viscosity axis; hence 
plotting stresses should prove more telling, close to a glass
transition, than plotting viscosities. The latter can reveal subtle
non-asymptotic corrections, but only if they are replotted on a smaller
scale as done in the insets of figures \ref{fig5}, \ref{fig6}. There,
various apparent power-law exponents could be fitted to the numerical
data. Considering the ISHSM as a model for colloidal suspensions, the
high-shear limiting viscosity contribution needs to be included;
neglecting hydrodynamic interactions, this is set by the solvent viscosity
$\eta_\infty$. The corresponding dot-dashed curves in figures \ref{fig4},
\ref{fig6}  show that a rather close approach to the critical packing
fraction is required for the structural stress captured in
\gl{iso4} to dominate. Considering that hydrodynamic interactions
cause an appreciable increase of the high-shear limiting viscosity
over the solvent one\cite{russel}, this condition appears severe.
However, it is obviously satisfied for systems that are nonergodic at
rest ($\varepsilon > 0$), if small enough $\gd$ can be achieved.

Above the transition, the quiescent system forms an (idealized)
glass\cite{Bengtzelius84,gs} 
which exhibits finite elastic constants. The
transverse elastic constant $G_\infty$ describes the (zero-frequency) Hookian
response of the amorphous solid to a small applied shear strain $\gamma$,
so that $\sigma = G_\infty \gamma$ for $\gamma\to0$. If steady flow is
imposed on the system, however, we find that the glass yields and is shear
melted by arbitrarily small shear rates. This fluidization is not
simply a trivial consequence of advection of the particles with the
flow (Taylor dispersion is included in neither model) but implies that
particles are freed from their cages and diffusion
perpendicular to the shear plane also becomes possible. Hence, within
our approach, infinitesimal steady shear leads to true melting of the
glass and not merely plastic flow of it. Any finite shear rate,
however small, 
sets a finite longest relaxation time, beyond which ergodicity is
restored; see the discussion of figures 
\ref{fig0}, \ref{fig1}.  

Nonetheless, a finite limiting stress (yield stress) must be overcome
in order to maintain the flow of the glass:
$\sigma(\gd,\eps>0)\to\sigma^+(\eps)$ for $\gd\to0$.  
To understand this better, note that
for $\varepsilon\ge0$ and $\gd\to0$, the time $\tau^{(\gd)}$
for the final decay , \gl{re1}, can become arbitrarily slow  compared
to the time 
characterizing the decay onto $f_q$.  
Inserting the scaling functions  $\Phi^{+}$ from
\glto{re2}{re5} into the expressions \gls{iso4}{sm3} for the stress,
 the long time contributions separate out. Importantly, the integrands
containing the $\Phi^{+}$ functions depend on time only via $\gd t$, 
so that nontrivial limits for the stationary stress follow in
the limit $\gd\to0$. In the
ISHSM for $\eps\ge0$,  this is given by (for $\gd>0$):
\beq{re8}
\sigma^+=\frac{k_BT\, \tilde{\gd}}{60\pi^2}\;
\int_0^\infty\!\! d\tilde t\, \int\!\! dk\; k^4 \;
\frac{S'_k S'_{\tilde{k}(\tilde t)}}{S^2_{\tilde k(\tilde t)}}\;
\left(\Phi^{+}_{k}(\tilde t)\right)^2 \; ,
\eeq
where $\tilde k(\tilde t) = k \left(1+ (\tilde{\gd} \tilde t)^2 / 3
\right)^{1/2}$, and the fixed reduced shear rate $\tilde{\gd}$ was
defined after \gl{re3}. The corresponding result in the
$F^{(\gd)}_{12}$ model is simply $\sigma^+=\tilde{\gd}\;
\int_0^\infty\!d\tilde t\; \Phi^+(\tilde t)$.

The existence of a dynamic yield stress in the glass phase is thus
 seen to arise
from the scaling law in \gls{re2}{re3}, whose  decay is initiated by
the shear-induced asymptote of the correlator $\cal G$ in \gl{re1}. Both
the ISHSM and the schematic model share this feature. The yield stress
arises from 
those fluctuations which require the presence of shearing to avoid
their arrest. Importantly, even though $\sigma^+$ requires the
solution of dynamical equations, within our approach it is solely and
uniquely determined 
by the equilibrium structure factor $S_q$ at the transition point. 
Assuming the connections of MCT to the potential energy paradigm for
glasses, as recently discussed \cite{Angelani00,Broderix00},
one might argue that $\sigma^+$ arises because the external
driving allows the system to overcome energy barriers so that
different metastable states can be reached. This interpretation 
would agree with the recent spin-glass\cite{Berthier00} and
soft-glassy rheology\cite{Sollich97,Sollich98b,Fielding00} approaches. 
Our microscopic approach indicates how shear achieves this in the case
of colloidal suspensions. It pushes 
fluctuations to shorter wavelengths where smaller particle
rearrangements cause their decay.

The increase of the amplitude of the yielding master functions
$\Phi^+$, see \gl{re4} originates in the increase of the arrested
structure in the unsheared glass (via the non-ergodicity parameters $f_q-f_q^c\propto
\sqrt\eps$). In consequence the
yield stress should rapidly increase as one moves further into the glass phase  $\sigma^+ -
\sigma^+_c\propto\sqrt{\eps}$ should (approximately) hold. Indeed,
the insets of  figures \ref{fig4}, \ref{fig5} show a good fit of this
square root increase to the numerical data. 
Our ability to make such a prediction is
highly significant: the yield stress is an important non-linear
property {\em of the arrested state itself}; it can be calculated
here, because the yield stress is governed by the onset of melting
under shear, which is itself a glass transition --- not at $\phi \to
\phi_c$ but at $\gd\to 0+$. 

An intriguing power-law increase of the stress above the yield value
at the critical point was noticed in Ref.\cite{Fuchs02} and is also
included (as dotted lines) in figures \ref{fig3} and \ref{fig4}. It
results from the leading corrections to the yielding scaling law
summarized in \gls{re6}{re7}.  Inserting those expressions into the
integrals for the stress leads to the small shear rate expansion
\beq{re9}
\sigma= \sigma^+ + |\gd t_0|^m \; \bar{\sigma} + {\cal O}( |\gd
t_0|^{2m}) \; ,\quad\mbox{for }\; |\eps| \ll |\gd t_0|^{\frac{2a}{1+a}}\; ,
\eeq
where the constant $\bar{\sigma}$ is given by an identical integral
to that in \gl{re8} but with $\Phi^+$ replaced with the correction
$\bar{\Phi}^+$.  For the ISHSM the exponent $m$ turns out as $m=0.152$
while it is $m=0.143$ for the $F^{(\gd)}_{12}$ model. 
As this prediction is based on the universal stability equation of a yielding 
glass, \gls{mi2}{mi3}, and on the existence of the yielding scaling
law, \gls{re2}{re3}, our approach suggest that this
Hershel-Bulkeley\cite{larson} flow curve may hold universally at a
glass transition point, with the exponent $m$ depending on the material
via the static structure factor $S_q$.

Another prediction from our asymptotic analysis of the leading
corrections is borne out in the numerical calculations. Deep in the
glass, \gls{re6}{re7} and (\ref{re9}) state that the leading
correction to $\sigma$ is independent on the shear rate $\gd$:
$\sigma=\sigma^++\eps^{m'}+{\cal O}(\eps^{2m'})$. This explains why in
figures \ref{fig3} and \ref{fig4} the stress starts to rise with $\gd$
appreciably only when Pe$_0$ becomes non-negligible. 

For the ISHSM, a caveat concerning the numerical results is
required for shear rates beyond about Pe$_0>10^{-2}$. 
The wavevector integrals in \gl{iso2} do not converge properly for
large $k$, and the results thus depend on the  chosen cutoff, $k_{\rm
max}= 39.8/d$.  This is implied by the crosses included in figure
\ref{fig4}, which were calculated for $\phi=\phi_c$ but larger cutoff,
 $k_{\rm max}= 119.8/d$. While only small differences (between the
crosses and diamonds) remain for Pe$_0<10^{-2}$, for larger shear rates the more accurate
integrals lead to larger stresses. While this could be an artefact of
the ISHSM, figure \ref{fig3} shows that the time-dependent non-linear
shear modulus $g(t,\gd)$  decays rapidly for such large Peclet
numbers. Because its initial value, $g(t=0,\gd)$, is the instantaneous
modulus of the unsheared system, $g(t=0,\gd)=G(t=0)=\langle
\sigma_{xy}^2\rangle/V$, this depends strongly on the particle
interaction potential, and, for hard spheres should actually
diverge\cite{Lionberger94,Naegele98}. As changing 
$k_{\rm max}$ strongly influences $G(t=0)$, we presume that 
finite $k_{\rm max}$ effectively softens the particle repulsion. 
Only calculations for more general potentials can show whether this
indicates a strong dependence of the non-linear stress on the
steepness of the potential for not very small Peclet numbers.
Interestingly, the glassy yield stress does not exhibit this strong
dependence. It should however be noted that the ISHSM underestimates
the effects of shearing  as the 
ratio  $\sigma^+_c/G^c_\infty=0.33$, is overestimated
\protect\cite{russel,larson,Nommensen99}.

\subsection{Structural versus non-structural features}
\label{sectnonuniv}

\begin{centering}
\begin{figure}[h]
\centerline{\psfig{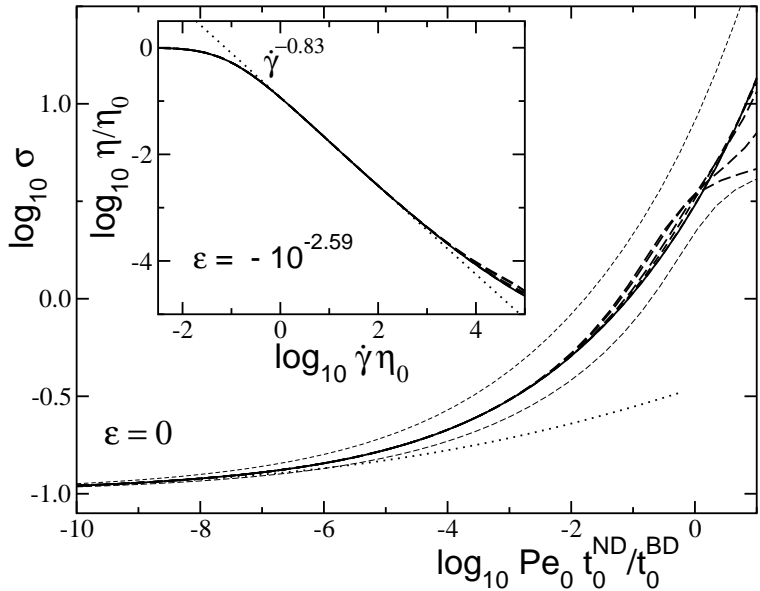}}  
\caption{In the main panel, steady state ``stress'',
$\sigma=\gd\langle \tau\rangle$, versus rescaled Pe$_0$ for the schematic
$F_{12}^{(\gd)}$ model with different  short time dynamics at the
critical point 
$\eps=0$; a dotted curve gives the Hershel-Bulkeley law from
figure \protect\ref{fig3}. In the inset,  shear rate dependent ``viscosity'',
$\eta=\langle \tau\rangle$,  for different short time dynamics 
below the transition, $\eps=-4^2\eps_1=-10^{-2.59}$, rescaled
with the shear rate independent $\eta_0=\eta(\gd=0)$; a dotted lines
gives a power-law fit. The viscosities $\eta_0$ at this $\eps$
also determine the rescaling factors used in the main panel:
$t^{\rm BD}_0/t^{\rm ND}_0=\eta^{\rm BD}_0/\eta^{\rm ND}_0$. 
Solid lines give
Brownian dynamics results, while dashed curves correspond to Newtonian
dynamics  with $\nu/\Omega=5$, 1 (both short), 0.5, 0.1 and
0.01 (all three long dashes). Thin lines in the main panel give the
unshifted $\sigma$ for  $\nu/\Omega=5$ and 0.01.}
\label{fig7}
\end{figure}
\end{centering}

Recently, computer simulations of sheared atomic glass
formers\cite{Yamamoto98,Barrat01,Berthier02}  have been
performed. While our microscopic theory is firmly based on assuming
colloidal dynamics, in the simplified models, the effects of a
different short time dynamics can easily be studied. Without shear,
the MCT has found that the long-time structural relaxation is
independent of the microscopic short time
dynamics\cite{Franosch98,Fuchs99}. (The latter only sets the overall
time scale via the matching time  $t_0$.)
In the F$_{12}^{(\gd)}$ model, in order to mimic Newtonian dynamics,
\gl{sm1} can be replaced by
\beq{sm5}
\ddot{\Phi}(t) + \nu \; \dot{\Phi}(t) + \Omega^2 \,  \left\{
\Phi(t) +  \int_0^t\!\!\! dt'\; m(t-t') \, \dot{\Phi}(t')
\right\} = 0 \; ,
\eeq
while \gl{sm2} remains. Here, $\Omega$ is a microscopic vibrational
frequency and $\nu$ a bare damping coefficient. Varying $\nu/\Omega$ shifts
$t_0$, and the thin lines in figure \ref{fig7} show the effect on the
stress of the critical glassy state. The factor $x=(t_0^{\rm
ND}\Omega)/(t_0^{\rm BD}\Gamma)$ varies  between 0.3 and 5.
As discussed in the context of the yielding scaling law,
\gls{re2}{re3} and \gl{re8}, the yield stress $\sigma^+$ is, according to our approach, a purely structural
property which is independent of the microscopic transient
dynamics. Beyond the limit $\gd\to0$, the one microscopic
matching time $t_0$ enters the structural dynamics and affects the
rise of $\sigma$. The data can be collapsed over a substantial
window by replotting them versus $\gd t_0$ as explained by
\gl{re9}; see the bold lines in figure \ref{fig7}. 
This suggests that the expansion \gl{re9} could be extended
to higher orders and might then obtain a larger range of validity. As
the average relaxation time $\tau_0=\langle \tau(\gd=0)\rangle$ of an
unsheared fluid state close to the transition 
is proportional to $t_0$, replotting the rescaled relaxation times
(``viscosities''),  $\langle \tau(\gd)\rangle/\tau_0$, versus rescaled
shear rate, $\gd\tau_0$, eliminates the $t_0$ dependence and leads to
data collapse even though the the expected asymptotic power law,
$\eta\propto\gd^{-1}$ is still strongly distorted.

To conclude, as argued already in section \ref{consteq}, even under shearing
the structural relaxation remains independent of short time microscopic 
effects, except for shifts in the matching time $t_0$.
According to this, simulations with either Newtonian or Brownian
dynamics (that are otherwise identical) should thus observe the same nonlinear rheology near the glass transition.   
Rescaling with $t_0$ eliminates transient features and the
nonlinear stress is only determined by the static structure factor for
Pe$_0\ll1$. This last statement bears an important caveat though. 
The time $t_0$ is rather fast and thus not well separated from other
microscopic time scales, so that corrections from finite Pe$_0$ need
to be anticipated. This remains a task for future work. 

\subsection{Non-linear Maxwell model}

In 1863, Maxwell suggested a simple model for the linear
rheological properties of a viscoelastic fluid, which has been the
cornerstone of the phenomenology of glassy systems since then. 
He suggested that the time dependent shear modulus decays exponentially,
$G(t)=G_\infty\, \exp{-t/\tau}$, where $G_\infty$ is the solid shear
modulus and $\tau$ the structural relaxation time. Viscosity and
stress follow as usual, $\sigma=\eta\gd=\gd\int_0^\infty\!\! dt\;
G(t)$. Our results suggest, as a toy model, a non-linear extension of
Maxwell's model which captures the competition of structural arrest
and shear-induced motion. It is obtained simply by postulating that
the transient shear modulus can relax via two independent processes, an
internal one characterized by $\tau$, and an induced or driven one
characterized by $\tau^{(\gd)}=c_*/|\gd|$ (with $c_*$ some numerical
constant): 
\beq{max1} 
g^{\rm n-l M}(t,\gd) = G_\infty \; e^{-t/\tau^{\rm M}}\; ,
\qquad\mbox{with }\; \frac{1}{\tau^{\rm M}} = \frac{1}{\tau} +
\frac{1}{\tau^{(\gd)}}\; .
\eeq
From this the expression for the flow curve follows immediately as
\beq{max2}
\sigma = \gd \; \left[ \eta_\infty +   \frac{G_\infty \tau}{1+|\gd|
\tau/c_*} \right] \; ,
\eeq
which exhibits a simplified scenario of the yielding-to flow
transition which we derived within our microscopic approach; a glass
would correspond to $1/\tau=0$, in which case the flow curve is that of a Bingham plastic. That toy model is not totally trivial, in that it gives 
a non-analytic dependence on the shear rate.

\section{Conclusions and outlook}

Building on Ref.\cite{Fuchs02}, we have presented a microscopic theory of the
nonlinear rheology of colloidal fluids and glasses under steady
shear. This predicts a universal transition between shear-thinning
fluid flow, with diverging viscosity upon increasing the interactions,
and solid yielding, with a yield stress that is finite at (and beyond)
the glass point.  Numerical calculations presented here
within progressively more simplified models support these universal
predictions. The novel yielding behaviour is seen to arise from a
competition between a collective caging of particles driven by
increasing local order, and shear advection of the involved density
fluctuations  to smaller wavelengths, where Brownian particle motion
relaxes them more effectively.

The shear-melting scenario can be rationalized solely from the stability
analysis of a yielding solid summarized by \gls{mi2}{mi3}.
As we tried to work out carefully, the properties of the correlator
$\cal G$ and some (natural but as yet unproven, except for the models
discussed here) assumptions about its extension to longer times
suffice to explain our major predictions. These include:
($i$) The existence of a dynamic yield stress of glasses, which increases
strongly with increasing density beyond the transition. 
($ii$) The relatively small increase of the potential part of the 
stress in the solid above the yield value upon increasing $\gd$. 
($iii$) The prediction of a Hershel-Bulkeley flow curve at the critical point, with a power law index fixed by the quiescent static structure factor.
($iv$) An asymptotic power-law shear thinning viscosity, $\eta\propto
\gd^{-1}$ in the fluid (this follows from the existence of a yield stress).
($v$) The observation that motions for times shorter than the final
relaxation time $\tau$ influence the stress or viscosity beyond
Pe $=\gd\tau\ll1$. This finding, discussed for a schematic model
in section \ref{sectnonuniv}, predicts strong sub-leading corrections to points ($ii$)-($iv$) above and shows that more 
detailed calculations (which ought to include hydrodynamic interactions)
are required, especially when no clear separation of time scales is present.

Reassuringly, the central non-linear stability equation, \gl{mi3},
which lies at the core of our predictions, can be argued on more
general grounds than the explicit derivation we gave. Without
shear, its predictions have been tested experimentally
\cite{Megen93,Megen94b,Goetze99}, and its structure thus appears
established (with some confidence certainly for $\eps<0$). The simplest
inclusion of shear naturally leads to a term $-(\gd t)^2$ as the essential
perturbation for long times. (Obviously the sign of $\gd$ must not
matter, and the term must be negative to induce relaxation.)
Also, because \gl{mi3} without shear is a
scaling equation without inherent time-scale (viz. ${\cal
G}(y*t,\gd=0)$ is a solution for any $y>0$ if it is for $y=1$), 
the shear-rate $\gd$ can only enter multiplied by time $t$
itself. Thus we expect \gl{mi3} to hold within a wider context
of non-ergodicity transitions and hope that the
predictions ($i$) to ($v$) above may apply to a broader class of scenarios describing yielding of metastable solids and shear thinning of 
viscoelastic fluids. 

On the other hand, the approach we have outlined cannot address the shear thickening behaviour that, for many colloidal materials, occurs at higher flow rates than those addressed here (but in some cases appears to preclude shear thinning altogether). It has been suggested \cite{Liu98,Head01} that this jamming phenomenon is connected with a {\em stress-induced} glass transition. This avenue will be explored in future work on a version of the schematic model in which \gl{sm2} is modified to include explicit stress- (as well as strain-rate-) dependence \cite{Holmes02}.

\begin{acknowledgments}
We thank J.-L. Barrat, J. Bergenholtz, L. Berthier,
A. Latz, K. Kroy and G. Petekidis for discussions. 
M.F.\ was supported by the DFG, grant Fu~309/3. 
\end{acknowledgments}

\appendix
\section{Bifurcation analysis of the microscopic
approach}

To determine steady state quantities like the stress, the
transient density fluctuations must be found within the suggested
microscopic approximation scheme. 
With $t$ denoting the time after switching on the shear field, the
normalized density correlators obey the exact equation of motion: 
\beq{a1}
\dot{\Phi}_\vek{q}(t) + \Gamma_\vek{q}(t)  \left\{
\Phi_\vek{q}(t) + 
 \int_0^t\!\!\! dt'\; m_\vek{q}(t,t') \,
\dot{\Phi}_\vek{q}(t')
\right\} = 0 \; ,
\eeq
where the ``initial decay rate'' becomes time-dependent with shear and
exhibits Taylor dispersion,
$\Gamma_\vek{q}(t)\, S_q = D_0\left( q^2+q_x q_y \gd t+( 
q_x q_y \gd t+q_x^2 \gd^2 t^2 ) S_q-(q_x q_y/q) \gd S'_q\right)$.
The memory function $m_\vek{q}(t,t')$ can only be found
after approximations, and, in the MCT spirit of our approach, is
approximated by projecting the fluctuating forces onto density pairs
and factorizing the resulting pair-density correlation functions:
\beq{a2}
 m_\vek{q}(t,t') \approx 
\frac{1}{2N} \sum_{\vek{k}}  V^{(\gd)}_{\vek{q},\vek{k}}(t,t')\;
 \Phi_{\vek{k}}(t-t') \, \Phi_{\vek{q}-\vek{k}}(t-t') \; .
\eeq
The vertex $ V^{(\gd)}$
is evaluated in the limit Pe$_0\ll1$  but for
large times so that  $\gd t$ and $\gd t'$ are  finite.
As $\gd\to 0$, it reduces to the standard MCT vertex
\cite{Bengtzelius84} and like the latter is uniquely determined by the
equilibrium structure factor, $S_q$. For long times, it vanishes.

The (approximate) microscopic equations contain the central bifurcation
singularity of MCT whose universal properties can be obtained
from a stability analysis close to the singularity. It is these
universal properties which the simplified models share.  
Around a quasi-arrested glassy structure, characterized by parameters $f^c_q$, the
\gls{a1}{a2} can be solved with the ansatz 
\beq{a3}
 \Phi_{\vek{q}}(t) =  f^c_{q} + (1- f^c_{q})^2
 {\cal G}_{\vek{q}}(t) \; ,
\eeq
where we use the fact that without shear the system is isotropic and the
 (critical) glass
form factor  $f^c_q$ therefore depends on the magnitude $q$ only. 
Splitting the vertex close to the transition into its critical value
 and deviations,
\beq{a4}
  V^{(\gd)}_{\vek{q},\vek{k}}(t,t')= V^{(0)c}_{\vek{q},\vek{k}}
+  \delta V^{(0)}_{\vek{q},\vek{k}} +
\Delta  V^{(\gd)}_{\vek{q},\vek{k}}(t,t')\;
\eeq
the stability of  $f^c_q$ can be determined.
Introducing the (standard)  abbreviations for time independent values,
$m^c_q =
\frac{1}{2N} \sum_{\vek{k}}  V^{(0)c}_{\vek{q},\vek{k}} f^c_k  f^c_p$ and
 $\varepsilon_q = \frac{1}{2N} \sum_{\vek{k}}  \delta
V^{(0)}_{\vek{q},\vek{k}} f^c_k  f^c_p$, and for matrix 
coefficients in the expansion in $\cal G$, 
$C^c_{qk} =
\frac{1}{N} \sum_{\vek{k'}}  V^{(0)c}_{\vek{q},\vek{k'}} f^c_k (1-
f^c_p)^2 \delta_{k,k'}$
and
$C^c_{qkp} =
\frac{1}{2N} \sum_{\vek{k'}}
 V^{(0)c}_{\vek{q},\vek{k'}} (1-f^c_k)^2 (1- f^c_p)^2\delta_{k,k'}
\delta_{p,|\vek{q}-\vek{k'}|}$, \gl{a1} for long times becomes
\beqa{a5}
& \frac{f^c_q}{(1-f^c_q)} - m^c_q +  \left[ {\cal G}_q(t)
- \sum_k C^c_{qk} {\cal G}_k(t) \right] - &
\nonumber \\&  \varepsilon_q - 
\frac{1}{2N} \sum_{\vek{k}}
\Delta  V^{(\gd)}_{\vek{q},\vek{k}}(t,0)\; f^c_k f^c_p - & \nonumber\\&
 \sum_{kp} C^c_{qkp} {\cal G}_k(t){\cal G}_p(t) +
\frac{d}{dt} \sum_{k} C^c_{qk} (1-f^c_q) \int_0^t\!\! dt'\;
{\cal G}_k(t-t'){\cal G}_q(t') + \ldots \; .& 
\eeqa
Here, a joint factor $(1-f^c_q)$ was divided out and ($\ldots$) denotes
both terms of higher order in $\cal G$ and those connected to the short
time dynamics which become negligible for long times. 
For large enough vertices, the first two terms in \gl{a5} of order
${\cal G}^0$ have a finite solution $f^c_q$, which defines  (idealized)
glass points. A critical (glass) point, viz. bifurcation singularity
in \gls{a1}{a2}, lies where the square bracket in \gl{a5} vanishes
because $C^c_{qk}$ has eigenvalue 1, and the linearized equation cannot
be solved for ${\cal G}(t)$, which gives the dynamics along the
unstable direction, ${\cal G}_q(t)={\cal G}(t) e_q$. Here the
eigenvectors of $C^c$ are denoted as $e_q$ and $\hat
e_q$, which satisfy $\sum_k C^c_{qk} e_k=e_q$ and $\sum_q \hat  e_q
C^c_{qk} =\hat e_k$ and the (conveniently chosen) normalizations
$\sum_q \hat e_q e_q = \sum_q \hat e_q (1-f^c_q) e_q e_q=1$. The
solvability equation for \gl{a5} results from requiring the contributions in the second and
third lines to be perpendicular to $\hat e_q$, and without shear
\gl{mi3} follows upon
the definitions: $\varepsilon=\sum_q \hat e_q \varepsilon_q$ and
$\lambda=\sum_{qkp} \hat e_q C^c_{qkp} e_k e_p$; for more details see
\cite{Goetze85,Goetze91b,Franosch97}. With shear, a time dependent
contribution arises from $\Delta  V^{(\gd)}_{\vek{q},\vek{k}}(t,0)$,
and because of spatial isotropy in the limit of small shear, this reduces
to:
\beq{a6}
t^2 \; c^{(\gd)} = \lim_{\gd\to0} \frac{-1}{2N\gd^2} \;
\sum_{\vek{q}\vek{k}}\; \hat e_{\vek{q}} \;
\Delta  V^{(\gd)}_{\vek{q},\vek{k}}(t,0)\; f^c_{|\vek{k}|}\,
 f^c_{|\vek{q}-\vek{k}|}\; ,
\eeq
where $\hat e_{\vek{q}}= \frac{\hat e_q}{4\pi q^2V}$ takes care of
three-dimensional integration. For colloidal hard spheres described with
the Percus-Yevick structure factor \cite{russel}, 
Monte Carlo integration  estimates give $ c^{(\gd)}\approx3$.

\section{Analysis of scaling equation (3)} 

The so-called $\beta$-correlator  $\cal G$ is the solution of \gl{mi3}
with the prescribed short time behavior ${\cal G}(t\to0) = (t_0/t)^a$
where $t_0$ is a matching time to the short time transient (determined
by $\Gamma$), and the ``critical'' exponent $a$ obeys
$\lambda=\Gamma^2(1-a)/\Gamma(1-2a)$. The derivation of this behaviour,
which solves \gl{a5} at the singularity for long times, and the
properties of $\cal G$ without shear can be found in Refs.\cite{Goetze85,Goetze91b,Franosch97}. With shear,
the analysis of \gl{mi3} resembles the one of the extended MCT
\cite{Fuchs92},
where a relaxation kernel describing thermally activated hopping
motion lead to a term linear in $t$ instead of the quadratic $\gd^2t^2$. 
Because, in the main text, we focus on the yielding behavior of
glassy solids, a discussion of the consequences of this quadratic term
close to the glass transition and in the glass shall now be given for the regime where shear-induced motion dominates.

A useful aspect of ${\cal G}$ are its homogeneity properties. With
$\Omega>0$  an arbitrary scale, it obeys
\beq{b1}
{\cal G}(t,\varepsilon,\gd) = \Omega^a \; \hat{\cal G}\left(\hat
t=\Omega(t/t_0), \hat \varepsilon=\varepsilon \Omega^{-2a},\hat{\gd}=\gd t_0
\Omega^{-(1+a)} \right)\; ,
\eeq
and thus does not change shape if the two control parameters are
varied on a scaling line $\gd t_0 = \hat{\gd} (\varepsilon/\hat
\varepsilon)^{(1+a)/(2a)}$. Moreover, this property enables one to
define three regimes, $(i)$  $\varepsilon \ll - \varepsilon_{\gd}$
far below the transition in the fluid region, where finite shear
barely distorts the fluid $\beta$-correlator;   
$(ii)$  $|\varepsilon| \ll  \varepsilon_{\gd}$, in the transition
region, where  the dynamical anomalies are most pronounced; and 
$(iii)$  $\varepsilon \gg  \varepsilon_{\gd}$ far above the transition
in the shear yielding glassy state. Here, $\varepsilon_{\gd}=
|\gd t_0|^{2a/(1+a)}$ defines a natural scale for the separation
parameter $\varepsilon$, and throughout the following, $|\varepsilon|\ll1$ and $\gd t_0
\ll1$ is assumed.

Equation (\ref{mi3}) can then be solved by power series expansions with
the general ansatz
\beq{b2}
{\cal G} = t^u \; \sum_{n=0} \alpha_n t^{vn}\; ,
\eeq
which yields
\beq{b3}
\varepsilon - c^{(\gd)} \; (\gd\; t)^2 =
\alpha_0^2\Gamma_{0,0} t^{2u} + 2 \alpha_0\alpha_1 \Gamma_{0,1}
t^{2u+v} + \sum_{n=2}t^{2u+nv} \sum_{n'=0}^n \Gamma_{n-n',n'}
\alpha_{n-n'}\alpha_{n'} \; .
\eeq
Here, an abbreviation employing Gamma functions $\Gamma(x)$ is used:
\beq{b4}
\Gamma_{n,n'}=
\frac{\Gamma{(1+u+nv)}\Gamma{(1+u+n'v)}}{\Gamma{(1+2u+(n+n')v)}} -
\lambda \; .
\eeq

\subsection{Glass for intermediate times}

For any $\gd$, a solution to \gl{b3} can be found for $\eps>0$ by
choosing $u=0$. This requires $\alpha_0=\sqrt{\eps/(1-\lambda)}$, 
$v=2$ and $\alpha_1=- c^{(\gd)}\gd^2/(2\alpha_0(1-\lambda))$, while the
higher coefficients can be found from a simple recursion relation. The
result (with $\Omega=\eps^{1/(2a)}$) takes the form
\beq{b5}
{\cal G}(\eps>0,t_{\rm f}\ll t\ll t_b) = 
\sqrt{\eps}\; 
\hat{\cal G}\left(\frac{t}{t_0}\eps^{\frac{1}{2a}},+1,\gd t_0
\eps^{-\frac{1+a}{2a}}  
\right) = \sqrt{\eps}\; 
\check{\cal G}( t/t_b ) =
\sqrt{\frac{\eps}{1-\lambda}} \; \left[ 1 - \frac{c^{(\gd)}}{2}
(\frac{t}{t_b})^2 + \ldots \right]\; ,
\eeq
where $t_b=\sqrt{\eps}/|\gd|$ is a time scale that
follows from balancing the first two terms in \gl{mi3}.
Without shear, the correlator would adhere to
$\sqrt{\eps/(1-\lambda)}$ for long times in the glass, while for
finite $\gd$ it relaxes below this value. The series \gl{b5} gives the
solution of \gl{mi3} for a finite window in time only, $t_{\rm f}\ll
t\ll t_b$, as can be seen
from its failure to match to the $(t_0/t)^{a}$ at short times, and as
it has a finite radius of convergence of the order of $t_b$; the
latter property can easily be determined from the recursion
relation for the $\alpha_{n\ge2}$. 
The obtained series thus describes, to order $\sqrt{\eps}$,
 the initial stage of the final  decay down from
$f_q=f^c_q+\sqrt{\eps/(1-\lambda)}$. 

\subsection{Transition region for long times}

If, close to the transition, $\eps=0$ can be set (region $(ii)$),
 then the choice $u=1$
requires $\alpha_0=\sqrt{c^{(\gd)}/(\lambda-\frac12)}\;
|\gd|$ in 
\gl{b3}. To match the resulting long-time series, higher
$\alpha_n$ are needed, and can be recursively be found, iff
$\Gamma_{0,1}$ vanishes:
\beq{b6}
\Gamma_{0,1}=\frac{\Gamma{(2)}\Gamma{(2+v)}}{\Gamma{(3+v)}}-\lambda =
0
\qquad \Leftrightarrow v=-(2-\frac1\lambda)\; .
\eeq
Obviously, coefficient $\alpha_1$ then cannot be determined from
\gl{b3}, but needs to be found from matching to ${\cal G}$ at shorter
times. Its scaling with $\gd$ can easily be obtained from the
homogeneity statement, \gl{b1}: $\alpha_1=\alpha_0\hat{\alpha}_1\,
t_{\gd}^{-1/v}$, where $\hat{\alpha}_1$ is a matching constant
(presumably of order unity), and the shear-rate dependent
$\beta$-scaling time appears
\beq{b7}
t_{\gd} = t_0 \; |\gd t_0|^{-\frac{1}{1+a}} \;. 
\eeq
As announced previously, the scaling function $\cal G$ then follows (with $\Omega=
|\gd t_0|^{\frac{1}{1+a}}$): 
\beq{b8}
{\cal G}(\eps=0,t\gg t_{\gd}) = 
\sqrt{\eps_{\gd}}\;
\hat{\cal G}\left(\frac{t}{t_{\gd}},0,1 \right) = 
 |\gd t_0|^{\frac{a}{1+a}}\;  \breve{\cal G}( t/t_{\gd} ) =
- \sqrt{\frac{c^{(\gd)}}{\lambda-\frac12}} \; |\gd|t\; \left[ 1 - \hat\alpha_1
(\frac{t}{t_{\gd}})^{-\frac{2\lambda-1}{\lambda}} + \ldots \right]\; .
\eeq
It matches to the decay of $\Phi(t)$ from unity down to $f^c$, which determines
$\hat\alpha_1$, and describes the initial stage of the shear-induced
decay of the correlators down from $f^c$ to zero. 

\subsection{Glass for long times}

Deep in the glass, region $(iii)$, the same reasoning as in the case
just studied can be used to determine $\alpha_1$ from matching --- now to
the series in \gl{b5}. With $u=1$ and $v$ from \gl{b6}, the scaling of
the undetermined coefficient is fixed by requiring it to extend
\gl{b5} from around $t_b$ to longer times:
$\alpha_1=\alpha_0\hat{\alpha}'_1\,
t_b^{-1/v}$, where $\hat{\alpha}'_1$ is another matching constant.
The resulting series continues the shear induced decay of the correlators
from the plateau value, as follows:
\beq{b9}
{\cal G}(\eps>0,t\gg t_b) = 
|\gd t_0|^{\frac{a}{1+a}}\; 
\hat{\cal G}\left(\frac{t}{t_{\gd}} ,\frac{\eps}{\eps_{\gd}}, 1
\right) =   
 \sqrt{\eps}\;  \tilde{\cal G}( t/t_b ) =
- \sqrt{\frac{c^{(\gd)}}{\lambda-\frac12}} \; |\gd|t\; \left[ 1 - \hat\alpha'_1
(\frac{t}{t_b})^{-\frac{2\lambda-1}{\lambda}} + \ldots \right]\; .
\eeq


\end{document}